\documentclass[12pt]{article}
\usepackage[dvips]{graphics,color}
\usepackage{latexsym,amssymb}
\usepackage{amsmath,amsbsy}
\usepackage[all]{xy}
\usepackage{amsopn,amstext}
\usepackage{cite}
\usepackage{amssymb}
\usepackage{rotating}
\usepackage{amsmath}
\usepackage{amsopn,amstext}
\allowdisplaybreaks[4]

\renewcommand{\theequation}{\arabic{section}.\arabic{equation}}

\begin{document}

\def\bea{\begin{eqnarray}}
\def\eea{\end{eqnarray}}
\def\no{\nonumber}

\baselineskip=20pt

\newcommand{\Title}[1]{{\baselineskip=26pt
   \begin{center} \Large \bf #1 \\ \ \\ \end{center}}}
\newcommand{\Author}{\begin{center}
   \large \bf
Guang-Liang Li${}^{a,b}$, Junpeng Cao${}^{b,c,d,e}\footnote{Corresponding author:
junpengcao@iphy.ac.cn}$, Wen-Li
Yang${}^{b,f,g}\footnote{Corresponding author:
wlyang@nwu.edu.cn}$, Kangjie Shi${}^{f,g}$ and Yupeng
Wang${}^{b,c}$
 \end{center}}

\newcommand{\Address}{\begin{center}
${}^a$ Ministry of Education Key Laboratory for Nonequilibrium
Synthesis and Modulation of Condensed Matter, School of
Physics, Xi'an Jiaotong University, Xi'an 710049, China\\
${}^b$ Peng Huanwu Center for Fundamental Theory, Xi'an 710127, China\\
${}^c$ Beijing National Laboratory for Condensed Matter Physics, Institute of Physics, Chinese Academy of Sciences, Beijing 100190, China\\
${}^d$ School of Physical Sciences, University of Chinese Academy of Sciences, Beijing 100049, China\\
${}^e$ Songshan Lake Materials Laboratory, Dongguan, Guangdong 523808, China \\
${}^f$ Institute of Modern Physics, Northwest University, Xi'an 710127, China\\
${}^g$ Shaanxi Key Laboratory for Theoretical Physics Frontiers, Xi'an 710127, China
\end{center}}

\Title{Spectrum of the quantum integrable $D^{(2)}_2$ spin chain with generic boundary fields}

\Author

\Address \vspace{0.3truecm}
\begin{abstract}

Exact solution of  the quantum integrable $D^{(2)}_2$ spin chain with generic integrable boundary fields is constructed.
It is found that the transfer matrix of this model can be factorized as the product of
those of two open staggered anisotropic XXZ spin chains.
Based on this identity, the
eigenvalues and Bethe ansatz equations of the $D^{(2)}_2$ model are derived via off-diagonal Bethe ansatz.

\vspace{0.5truecm} \noindent {\it PACS:} 75.10.Pq, 02.30.Ik,
71.10.Pm

\end{abstract}
\newpage

\section{Introduction}
\label{intro} \setcounter{equation}{0}

The $D^{(2)}_2$ spin chain model is one of the most representative integrable system associated with  quantum algebra beyond $A$-series.
The exact solution of  the $D^{(2)}_2$ spin chain is also the foundation to solve the high rank $D^{(2)}_{n}$ models with nested analytical methods.
Particularly, the $D^{(2)}_2$ spin chain has many applications in the string theory and black hole.
For an example, Robertson, Jacobsen and Saleur found \cite{b1} that an open $D^{(2)}_2$ spin chain with some integrable boundary condition possesses the lattice regularisation of a
non-compact boundary conformal field theory and is closely related to the SL(2, $\mathbb{R}$)/U(1)
Euclidean black hole \cite{17,18,19,20,21}.

The eigenvalues of the transfer matrix of the periodic $D^{(2)}_n$ model firstly was obtained by the analytical Bethe ansatz\cite{25} and then by  the algebraic Bethe ansatz \cite{m1}.
For open boundary conditions, besides the $R$-matrix, the reflection matrices should also be used to construct
the transfer matrix which generates the conserved quantities including the model Hamiltonian \cite{op1,op2,op3}.
The Hamiltonian with diagonal boundary fields was exactly solved via both the coordinate Bethe ansatz \cite{26} and the analytical Bethe ansatz \cite{27, 28}.
Recently, Robertson, Pawelkiewicz, Jacobsen and Saleur \cite{b2} reported that the $R$-matrix of $D^{(2)}_2$ model \cite{13,14,15} is related to the
antiferromagnetic Potts model and the staggered XXZ spin chain \cite{4,5,6,7,8,9}.
Based on this idea, Nepomechie and Retore \cite{b5} obtained the exact solutions of transfer matrices of both the closed $D^{(2)}_2$ spin chain and the open one with a special boundary condition
by using the factorization identities and algebraic Bethe ansatz.

In this paper, we study the exact solution of the $D^{(2)}_2$ spin chain with generic non-diagonal boundary fields. Because the reflection matrix and the dual one can not be diagonalized simultaneously,
the $U(1)$ symmetry of the system is broken.
The structure of the present paper is as follows. In section 2, we give a brief description of the $D^{(2)}_2$ model with open boundary condition.
The $R$-matrix, reflection matrices and generating functional of conserved quantities
are introduced. In section 3, we show that the transfer matrix can be factorized as the product of two open staggered XXZ spin chains.
In section 4, by using the fusion techniques, we obtain the exact solution of the system via off-diagonal Bethe ansatz. The inhomogeneous $T-Q$ relations and related Bethe ansatz equations are given.
The summary of main results and some concluding remarks are presented in section 5. Appendix A  provides the results for another inequivalent generic non-diagonal boundary fields.

\section{$D^{(2)}_2$-model}
\setcounter{equation}{0}

The conserved quantities including the model Hamiltonian of the $D_2^{(2)}$ spin chain  are generated by the transfer matrix $t(u)$
\begin{equation}
t(u)= tr_0 \{K^+_0(u)T_0(u) K^-_0(u)\hat{T}_0 (u)\}. \label{trweweu1110}
\end{equation}
Here $u$ is the spectral parameter, the subscript $0$ means the four-dimensional auxiliary space $V_0$,
$tr_0$ means taking trace only in the auxiliary space $V_0$, $K^+_0(u)$ is the boundary reflection matrix defined in the auxiliary space at one end,
$K^-_0(u)$ is the reflection matrix at the other end,
$T_0(u)$ and $\hat{T}_0(u)$ are the monodromy matrices constructed by the
$16\times 16$ $R$-matrix as
\bea
&& T_0(u)=R_{01}(u)R_{02}(u)\cdots R_{0N}(u), \no\\
&&\hat{T}_0 (u)=R_{N0}(u)R_{N-10}(u)\cdots R_{10}(u).\label{Tt11}
\eea
Here the subscript $j=1, \cdots, N$ denotes the four-dimensional quantum space $V_j$ of $j$-th site, which means that
the spin of the $D_2^{(2)}$ chain at $j$-th site has four components, and $N$ is the number of sites.
Thus $T_0(u)$ and $\hat{T}_0(u)$ are defined in the tensor space $V_0 \otimes V_1\otimes  \cdots \otimes V_{N}$ and
$\otimes_{j=1}^N V_{j}$ is the quantum or physical space.

The integrability of the system requires that the transfer matrices \eqref{trweweu1110} with different spectral parameters commutate with each other
\begin{equation}
[t(u), t(v)]=0. \label{20211224-1}
\end{equation}
Thus all the expansion coefficients of $t(u)$ with respect to $u$ are commutative.
The coefficients or their combinations are the conserved quantities.
The commutation relation \eqref{20211224-1} is achieved by that the
$R$-matrices in Eq.\eqref{Tt11} satisfy the Yang-Baxter equation
\begin{eqnarray}
R_{12}(u-v)R_{13}(u)R_{23}(v)=R_{23}(v)R_{13}(u)R_{12}(u-v), \label{20190802-1}
\end{eqnarray}
and the reflection matrices in Eq.\eqref{trweweu1110} for the
given $R$-matrix satisfy the reflection equations\cite{op1,op2,op3}
\begin{eqnarray}
&&R_{12}(-u+v){K^+_{1}}(u)M_1^{-1}R_{21}
 (-u-v+4\eta)M_1{{K}^+_{2}}(v)\nonumber\\
&&\qquad={{K}^+_{2}}(v)M_1R_{12}(-u-v+4\eta)M_1^{-1} K^+_{1}(u)R_{21}(-u+v),
 \label{r2}\\[6pt]
&&  R_{12}(u-v)K^-_{1}(u)R_{21}(u+v) {K^-_{2}}(v)=
 K^-_{2}(v)R_{12}(u+v){K^-_{1}}(u)R_{21}(u-v). \label{r1}
\end{eqnarray}

The solution of Yang-Baxter equation \eqref{20190802-1} associated with  the twisted $D^{(2)}_2$ quantum
algebra,  gives the $16\times 16$ $R$-matrix defined in the tensor space
$V_1 \otimes V_2$ as \cite{26,27,28}
\bea
&&R_{12}(u)=e^{-2(u+2\eta)}\Big\{
(e^{2u}-e^{4\eta})(e^{2u}-e^{4\eta})\sum_{\alpha\neq
2,3}[e_1]^{\alpha}_{\alpha}\otimes [e_2]^{\alpha}_{\alpha}+e^{2\eta}(e^{2u}-1)(e^{2u}-e^{4\eta}) \nonumber\\[4pt]
&&\qquad \times
\sum_{\substack{\alpha\neq\beta,\beta'\\\alpha\,\textrm{or}\,\beta\neq
2,3}} [e_1]^{\alpha}_{\alpha}\otimes [e_2]^{\beta}_{\beta} -\frac{1}{2} (e^{4\eta}-1)(e^{2u}-e^{4\eta}) \Big[ (e^{u}+1)
\Big(\sum_{\alpha=1,\beta=2,3}+e^{u}\sum_{\alpha=4,\beta=2,3} \Big) \no\\[4pt]
&&\qquad \times \Big( [e_1]^{\alpha}_{\beta}\otimes
[e_2]^{\beta}_{\alpha}+[e_1]^{\beta'}_{\alpha'}\otimes [e_2]^{\alpha'}_{\beta'} \Big)
+ (e^{u}-1)
\Big(-\sum_{\alpha=1,\beta=2,3}+e^{u}\sum_{\alpha=4,\beta=2,3} \Big)
 \no\\[4pt]
&&\qquad \times \Big( [e_1]^{\alpha}_{\beta}\otimes
[e_2]^{\beta'}_{\alpha}+[e_1]^{\beta'}_{\alpha'}\otimes [e_2]^{\alpha'}_{\beta}
\Big)\Big]+ \sum_{\alpha,\beta\neq
2,3}a_{\alpha\beta}(u)[e_1]^{\alpha}_{\beta}\otimes [e_2]^{\alpha'}_{\beta'}+
 \frac{1}{2}\sum_{\alpha\neq 2,3,\beta=2,3}
 \no\\[4pt]
&&\qquad \times \Big[ b_\alpha^{+}(u) \Big( [e_1]^{\alpha}_{\beta}\otimes
[e_2]^{\alpha'}_{\beta'}+[e_1]^{\beta'}_{\alpha'}\otimes [e_2]^{\beta}_{\alpha}
 \Big)
+ b_\alpha^{-}(u) \Big( [e_1]^{\alpha}_{\beta}\otimes
[e_2]^{\alpha'}_{\beta}+[e_1]^{\beta}_{\alpha'}\otimes [e_2]^{\beta}_{\alpha}
 \Big)\Big]
 \no\\[4pt]
&&\qquad+\sum_{\alpha=2,3} \Big(c^{+}(u)[e_1]^{\alpha}_{\alpha}\otimes
[e_1]^{\alpha'}_{\alpha'}+ c^{-}(u)[e_1]^{\alpha}_{\alpha}\otimes
[e_2]^{\alpha}_{\alpha}+d^{+}(u)[e_1]^{\alpha}_{\alpha'}\otimes [e_2]^{\alpha'}_{\alpha}
\no\\[4pt]
&&\qquad +
d^{-}(u)[e_1]^{\alpha}_{\alpha'}\otimes [e_2]^{\alpha}_{\alpha'} \Big)\Big\}.
\label{RD2}
\eea
Here $\eta$ is the crossing parameter, $\alpha$ and $\beta$ take the values from 1 to 4, $\alpha'=5-\alpha$, $\beta'=5-\beta$,
$\bar{\alpha}=2$ if $\alpha =1$, $\bar{\alpha}=\frac{5}{2}$ if $\alpha=2$ or $\alpha=3$, and $\bar{\alpha}=3$ if $\alpha=4$.
$[e_k]^{\alpha}_{\beta}$ $ (k=1,2) $ is the $4\times 4$ representation matrix of Weyl basis of $k$-th space.
The coefficients $a_{\alpha\beta}(u)$  are defined as
\begin{equation}
a_{\alpha\beta}(u)=\begin{cases}
(e^{4\eta}e^{2u}-e^{4\eta})(e^{2u}-1), & \alpha=\beta, \\[4pt]
(e^{4\eta}-1)[e^{4\eta}e^{2\eta(\bar\alpha-\bar\beta)}(e^{2u}-1)-\delta_{\alpha\beta'}(e^{2u}-e^{4\eta})], & \alpha<\beta, \\[4pt]
(e^{4\eta}-1)e^{2u}[e^{2\eta(\bar\alpha-\bar\beta)}(e^{2u}-1)-\delta_{\alpha\beta'}(e^{2u}-e^{4\eta})], & \alpha>\beta,
\end{cases}
\end{equation}
where $\alpha,\beta\neq 2,3$. The functions $b_{\alpha}^{\pm}(u)$, $c^{\pm}(u)$ and $d^{\pm}(u)$ are given by
\bea
&&b_{\alpha}^{\pm}(u)=\begin{cases}
\pm e^{2\eta(\alpha-1/2)}(e^{4\eta}-1)(e^{2u}-1)(e^u\pm e^{2\eta}), & \alpha=1, \\[4pt]
e^{2\eta(\alpha-7/2)}(e^{4\eta}-1)(e^{2u}-1)e^u(e^u\pm e^{2\eta}), & \alpha=4,
\end{cases} \nonumber \\[4pt]
&&c^{\pm}(u)=\pm\frac{1}{2}(e^{4\eta}-1)(e^{2\eta}+1)e^u(e^u\mp
1)(e^u\pm e^{2\eta})+e^{2\eta}(e^{2u}-1)(e^{2u}-e^{4\eta}), \no \\[4pt]
&& d^{\pm}(u)=\pm\frac{1}{2}(e^{4\eta}-1)(e^{2\eta}-1)e^u(e^u\pm
1)(e^u\pm e^{2\eta}). \eea

The $R$-matrix \eqref{RD2} has following properties
\begin{eqnarray}
{\rm Unitarity}&:&R_{12}(u)\,R_{21}(-u)=\rho(u)=16\sinh^2(u-2\eta)\sinh^2(u+2\eta),\nonumber\\
{\rm Initial~ condition}&:&R_{12}(0)=\rho(0)^{\frac{1}{2}}{\cal P}_{12},\label{Initial-cond}\\
{\rm Crossing \,unitarity}&:&R_{12}(u)^{t_1}M_1R_{21}(-u+4\eta)^{t_1}M_1^{-1}=\rho(u-2\eta), \no\\
&&R_{12}(u)^{t_2}M_2^{-1}R_{21}(-u+4\eta)^{t_2}M_2=\rho(u-2\eta),\label{Crossing-Unitarity}
\end{eqnarray}
where ${\cal P}_{12}$ is the permutation operator with the matrix
elements $[{\cal P}_{12}]^{\alpha\gamma}_{\beta\delta}=\delta_{\alpha\delta}\delta_{\beta\gamma}$, $R_{21}(u)={\cal
P}_{12}R_{12}(u){\cal P}_{12}$, $t_{k}$ denotes the transposition in the $k$-th space, $M_k$ is the $4\times 4$ diagonal constant matrix \bea
M_k=diag(e^{2\eta},1,1,e^{-2\eta}).\eea

The solutions of reflection equations \eqref{r2}-\eqref{r1} with fixed $R$-matrix \eqref{RD2} give the reflection matrices
$K_k^{\pm}(u)$ defined in the four-dimensional space $V_k$ as \cite{5-2,5-3,5-5,5-6}
\bea
&&K_k^+(u)=M_k K_k^-(-u+2\eta)|_{\{s,s_1,s_2\}\rightarrow \{s',s'_1,s'_2\}}, \label{kM}\\[8pt]
&& K_k^-(u)=\left(\begin{array}{cccc}
    k_{11}(u)&k_{12}(u)&k_{13}(u)&k_{14}(u) \\
    k_{21}(u)&k_{22}(u)&k_{23}(u)&k_{24}(u) \\
    k_{31}(u)&k_{32}(u)&k_{33}(u)&k_{34}(u) \\
    k_{41}(u)&k_{42}(u)&k_{43}(u)&k_{44}(u)
\end{array}\right),\label{kd-2matrix}
\eea where $\{s, s_1, s_2\}$ are the free boundary parameters at
one end and $\{s', s'_1, s'_2\}$ are the ones at the other end.
Here we should note that the reflection equation \eqref{r1} has
two inequivalent classes of generic non-diagonal solutions. Without losing generality, we consider one of the generic solutions, whose matrix elements are\footnote{The solution is different from that
given in \cite{5-6} even it has the same number of free boundary parameters. }
\bea&&k_{11}(u)=\frac{1}{2}e^{-u}[\cosh(u-\eta)\sinh(u-2s)-2s_1s_2\sinh\eta\sinh^2(u)],\no\\
&&k_{12}(u)=\frac{1}{2}s_1e^{-\frac
u2}\sqrt{\cosh\eta}\sinh(2u)\cosh\frac
12(u-\eta-2s),\no\\
&&k_{13}(u)=-\frac{1}{2}s_1e^{-\frac
u2}\sqrt{\cosh\eta}\sinh(2u)\sinh\frac
12(u-\eta-2s),\no\\
&&k_{14}(u)=\frac{1}{2}s_1^2\sinh u\sinh(2u),\no\\
&&k_{21}(u)=\frac{1}{2}s_2e^{-\frac
u2}\sqrt{\cosh\eta}\sinh(2u)\cosh\frac
12(u-\eta-2s),\no\\
&&k_{22}(u)=-\frac{1}{2}\cosh u[\sinh u+\cosh\eta\sinh(2s)],\no\\
&&k_{23}(u)=-\frac{1}{2}\sinh u[\sinh\eta\cosh(2s)+2s_1s_2\sinh u\cosh(u-\eta)],\no\\
&&k_{24}(u)=-\frac{1}{2}s_1e^{\frac
u2}\sqrt{\cosh\eta}\sinh(2u)\sinh\frac
12(u-\eta+2s),\no\\
&&k_{31}(u)=-\frac{1}{2}s_2e^{-\frac
u2}\sqrt{\cosh\eta}\sinh(2u)\sinh\frac
12(u-\eta-2s),\no\\
&&k_{32}(u)=-\frac{1}{2}\sinh u[\sinh\eta\cosh(2s)+2s_1s_2\sinh(u)\cosh(u-\eta)],\no\\
&&k_{33}(u)=\frac{1}{2}\cosh u[\sinh u-\cosh\eta\sinh(2s)],\no\\
&&k_{34}(u)=-\frac{1}{2}s_1e^{\frac
u2}\sqrt{\cosh\eta}\sinh(2u)\cosh\frac
12(u-\eta+2s),\no\\
&&k_{41}(u)=\frac{1}{2}s_2^2\sinh u\sinh(2u),\no\\
&&k_{42}(u)=-\frac{1}{2}s_2e^{\frac
u2}\sqrt{\cosh\eta}\sinh(2u)\sinh\frac
12(u-\eta+2s),\no\\
&&k_{43}(u)=-\frac{1}{2}s_2e^{\frac
u2}\sqrt{\cosh\eta}\sinh(2u)\cosh\frac
12(u-\eta+2s),\no\\
&&k_{44}(u)=-\frac{1}{2}e^{u}[\cosh(u-\eta)\sinh(u+2s)-2s_1s_2\sinh\eta\sinh^2(u)].\label{kd21q3-1}
\eea For the $K$-matrices $K_k^{\pm}(u)$ given by (\ref{kd21q3-1}) and (\ref{kM}) satisfy $tr_0  K^{+}_0(0)\ne 0$. The Hamiltonian
can be given in terms of the transfer matrix by the standard way\footnote{It is remarked that one can define the
Hamiltonian by (\ref{hh}) for the case of $t(0)\ne 0$ (i.e., $tr_0
 K^{+}_0(0)\ne 0$), however for the case of $t(0)=0$  one needs to adopt other way \cite{b1} instead
to construct a meaningful Hamiltonian.} \cite{op2}
\begin{eqnarray}
H= \frac{\partial \ln t(u)}{2\partial
u}|_{u=0,\{\theta_j\}=0}-\frac{tr_0{{K}^+_0}(0)'}{2tr_0{K}^+_0(0)}
=\sum^{N-1}_{k=1}H_{k
k+1}+\frac{{K^{-}_N}(0)'}{2{K^{-}_N}(0)}+\frac{ tr_0 \{
K^{+}_0(0)H_{10}\}}{tr_0  K^{+}_0(0)}, \label{hh}
\end{eqnarray} where $H_{k\,k+1}= \rho(0)^{-1}{R}_{k\,k+1}(0)\,\frac{\partial}{\partial u}R_{k\,k+1}(u)\left.\right|_{u=0}.$

Another solution with 3 free boundary parameters is given by
\eqref{sskd-1} below, which agrees with that obtained in
\cite{5-6}. It
is easy to check that $K^+(u)$ and $K^-(u)$ can not be
diagonalized simultaneously for generic choices of 6 boundary
parameters. Then the traditional algebraic Bethe ansatz can
not be applied to solve the eigenvalues of transfer matrix
\eqref{trweweu1110} because of the absence of an obvious reference state
\cite{wang15}.

\section{Factorization of the reflection matrices}
\setcounter{equation}{0}

To obtain the eigenvalues of the transfer matrix
\eqref{trweweu1110}, we first consider the decomposition of space.
The four-dimensional space can be regarded as the tensor of two
two-dimensional spaces. For example, $V_1=V_{1'}\otimes V_{2'}$
and $V_2=V_{3'}\otimes V_{4'}$. Then the $R$-matrix \eqref{RD2}
can be factorized as the product of $R$-matrices of the
anisotropic XXZ spin chain with suitable global transformation
\cite{b1,b2,m2,b5} \bea && R_{12}(u)=2^4[S\otimes S]\tilde
R_{1'4'}(u+i\pi)\tilde R_{1'3'}(u)
\tilde R_{2'4'}(u)\tilde R_{2'3'}(u-i\pi) [S\otimes S]^{-1},\label{Factor-R-1}\\[4pt]
&&R_{21}(u)=2^4[S\otimes S]\tilde R_{3'2'}(u+i\pi)\tilde R_{4'2'}(u)
\tilde R_{3'1'}(u)\tilde R_{4'1'}(u-i\pi) [S\otimes S]^{-1},\label{Factor-R-2}\eea
where the transformation matrix $S$ is
\bea
 S=S^{-1}=\left(\begin{array}{cccc}
    1&&& \\
    &\frac{\cosh\frac \eta2}{\sqrt{\cosh\eta}}&-\frac{\sinh\frac \eta2}{\sqrt{\cosh\eta}}& \\[4pt]
    &-\frac{\sinh\frac \eta2}{\sqrt{\cosh\eta}}&-\frac{\cosh\frac \eta2}{\sqrt{\cosh\eta}}& \\[4pt]
    &&&1\end{array}\right), \label{s1-matrix}\eea
and the $R$-matrix reads
\bea
 \tilde R_{1'2'}(u)=\left(\begin{array}{cccc}
    \sinh(-\frac u2+\eta)&&&\\
    &\sinh \frac u2 &e^{-\frac u2}\sinh \eta &  \\[4pt]
    &e^{\frac u2}\sinh \eta &\sinh\frac u2 &  \\[4pt]
    &&&\sinh(-\frac u2+\eta)
           \end{array}\right).\label{Rs-matrix}
\eea
The $R$-matrix \eqref{Rs-matrix} has following properties
\begin{eqnarray}
{\rm Quasi-period}&:&\tilde R_{1'2'}(u+2i\pi)=-\tilde R_{1'2'}(u),\nonumber\\
{\rm PT-symmetry}&:&\tilde R^{t_{1'}t_{2'}}_{1'2'}(u)=\tilde R_{1'2'}(u),\nonumber\\
{\rm Unitarity}&:&\tilde R_{1'2'}(u)\,\tilde R_{2'1'}(-u)=\rho_s(u)=\sinh(-\frac u2+\eta)\sinh(\frac u2+\eta),\nonumber\\
{\rm Initial~ condition}&:&\tilde R_{1'2'}(0)=\rho_s(0)^{\frac{1}{2}}{\cal \tilde P}_{1'2'},\nonumber\\
{\rm Crossing \,unitarity}&:&\tilde R_{1'2'}(u)^{t_{1'}}\tilde{M}_{1'}\tilde R_{2'1'}(-u+4\eta)^{t_{1'}}\tilde{M}_{1'}^{-1}=\rho_s(u-2\eta),\label{Crossing-Unitarity-1}
\end{eqnarray}
where ${\cal \tilde P}_{1'2'}$ is the permutation operator defined in the tensor space $V_{1'}\otimes V_{2'}$, $t_{k'}$ $ (k'=1',2')$ denotes the transposition in the $k'$-th subspace, and
$\tilde{M}_{k'}$ is the diagonal constant matrix with the form of $ \tilde{M}_{k'}=diag(e^{\eta},e^{-\eta})$.
Besides, the $R$-matrix \eqref{Rs-matrix} also satisfies the Yang-Baxter equation
\begin{eqnarray}
\tilde R_{12}(u-v)\tilde R_{13}(u)\tilde R_{23}(v)=\tilde R_{23}(v)\tilde R_{13}(u)\tilde R_{12}(u-v). \label{Ttq11we11}
\end{eqnarray}

The very factorization (\ref{Factor-R-1})-(\ref{Factor-R-2}) of the $R$-matrices allows us, after a tedious calculation,  to have  that the reflection matrices \eqref{kM}-\eqref{kd-2matrix} with the elements \eqref{kd21q3-1} can be
expressed in terms of the factorization form as
\bea
&&{K}^+_1(u)=[\rho_s(i\pi)]^{-\frac12}S\tilde R_{2'1'}(i\pi)\tilde{K}^+_{2'}(u)\tilde{M}_{2'}^{-1}\tilde R_{1'2'}(-2u+4\eta-i\pi)
\tilde{M}_{2'}\tilde{K}^+_{1'}(u)S^{-1}, \no \\[4pt]
&&K^-_1(u)=[\rho_s(i\pi)]^{-\frac12}S \tilde K^-_{1'}(u+i\pi)\tilde R_{2'1'}(2u+i\pi)\tilde K_{2'}^-(u)\tilde R_{1'2'}(-i\pi)S^{-1}, \label{kd1}
\eea
where $\tilde K^{\pm}_{k'}(u)$ are the $2\times 2$ generic non-diagonal reflection matrices of the XXZ spin chain
\bea
&&\tilde K^+_{k'}(u)=\tilde {M}_{k'} \,\tilde K_{k'}^-(-u+2\eta)|_{\{s,s_1,s_2\} \rightarrow \{s',s'_1,s'_2\}}, \label{ksk111}\\[6pt]
&&\tilde K^-_{k'}(u)=\left(\begin{array}{cc}
    -e^{-\frac u2}\,\sinh(\frac u2-s)&s_1\,\sinh u \\[4pt]
   s_2\,\sinh u &e^{\frac u2}\,\sinh(\frac u2+s)
   \end{array}\right), \label{kewesk111}
\eea
which satisfy the reflection equations
\bea
\hspace{-1.2truecm}&&\hspace{-1.2truecm}\tilde R_{1'2'}(-u+v){\tilde {K}^+_{1'}}(u)\tilde {M}_{1'}^{-1}\tilde R_{2'1'}
 (-u-v+4\eta)\tilde {M}_{1'}{\tilde{K}^+_{2'}}(v)\nonumber\\[4pt]
\hspace{-1.2truecm}&&\hspace{-1.2truecm}\qquad\qquad\qquad\qquad ={\tilde {K}^+_{2'}}(v)\tilde{M}_{1'}\tilde R_{1'2'}(-u-v+4\eta)\tilde {M}_{1'}^{-1}
{{\tilde K}^+_{1'}}(u)\tilde R_{2'1'}(-u+v), \label{rs12}\\[4pt]
\hspace{-1.2truecm}&&\hspace{-1.2truecm}\tilde R_{1'2'}(u-v)\tilde  K_{ 1'}^-(u)\tilde R_{2'1'}(u+v) \tilde K_{2'}^-(v)=
 \tilde K_{2'}^-(v)\tilde R_{1'2'}(u+v) \tilde K_{1'}^-(u)\tilde R_{2'1'}(u-v). \label{rs1}
\eea

Some remarks are in order. The boundary parameters  $s,\,s_1\,s_2$ are the same as
those of (\ref{kM})-(\ref{kd-2matrix}). Here it should also be
addressed that when $\tilde K^-(u)=1$ in (\ref{kewesk111}), the resulting
$K^-(u)$ given by (\ref{kd1}) is just that discussed in
refernce\cite{b5} with $\epsilon=0$. When $s_1=s_2=0$ in
(\ref{kewesk111}), the resulting $K^-(u)$ given by (\ref{kd1}) is the second
case  discussed in \cite{26}.  Due to the fact that $\tilde{K}^{\pm}(u)$ are all diagonal ones\footnote{However, the resulting $K^{\pm}(u)$ obtained by the relation (\ref{kd1}) even have  non-diagonal matrix elements\cite{26,b5}. }, it is  only special cases that one can adopt coordinate/algebraic
Bethe ansatz to solve the corresponding $D^{(2)}_2$ model \cite{b2,b5}.

Based on the $R$-matrix \eqref{Rs-matrix} and reflection matrices \eqref{ksk111}-\eqref{kewesk111}, we construct
the transfer matrix $\tilde t(u)$ of the inhomogeneous XXZ spin chain as
\begin{equation}
\tilde t(u)= tr_{0'} \{\tilde K^+_{0'}(u)\tilde T_{0'} (u)\tilde K^-_{0'}(u)\hat{\tilde T}_{0'}(u)\}, \label{ts-1}
\end{equation}
where $0'$ means the auxiliary space, $\tilde T_{0'}(u)$ and $\hat{\tilde T}_{0'}(u)$ are the monodromy matrices
\bea
&& \tilde T_{0'}(u)=\tilde R_{0'1'}(u-\theta_1)\tilde R_{0'2'}(u-\theta_2)\cdots
\tilde R_{0'(2N)'}(u-\theta_{2N}), \no\\
&&\hat{\tilde T}_{0'} (u)=\tilde R_{0'(2N)'}(u+\theta_{2N})\tilde R_{0'(2N-1)'}(u+\theta_{2N-1})
\cdots \tilde R_{0'1'}(u+\theta_1),\label{Tt1221}
\eea
and $\{\theta_j|j=1, \cdots, 2N \}$ are the inhomogeneous parameters.
We should note that the quantum space of transfer matrix $\tilde t(u)$ for the XXZ spin chain and that of
$t(u)$ for the $D^{(2)}_2$ model are the same. Thus the number of sites in Eq.\eqref{Tt1221} is extended to $2N$ to ensure the dimension of Hilbert space is $4^N$.
The monodromy matrices \eqref{Tt1221} satisfy the Yang-Baxter relations
\bea
&& \tilde R_{0'0''}(u-v) \tilde T_{0'}(u)\tilde T_{0''}(v)=\tilde T_{0''}(v)\tilde T_{0'}(u)\tilde R_{0'0''}(u-v), \no\\
&& \tilde R_{0''0'}(u-v) \hat {\tilde T}_{0'}(u) \hat {\tilde T}_{0''}(v)=\hat {\tilde T}_{0''}(v)\hat {\tilde T}_{0''}(u)\tilde R_{0''0'}(u-v).\label{Ttqwe11}
\eea
From the reflection equations \eqref{rs12}-\eqref{rs1} and Yang-Baxter relation \eqref{Ttqwe11}, we can prove that
the transfer matrices $\tilde{t}(u)$ with different spectral parameters commutate with each other
\begin{equation}
[\tilde t(u), \tilde t(v)]=0. \label{120211224-1}
\end{equation}

Interestingly, we find that if the inhomogeneous parameters are staggered, i.e., $\theta_j=0$ for the odd $j$ and $\theta_j= i\pi$ for the even $j$,
the transfer matrix \eqref{trweweu1110} of the $D^{(2)}_2$ spin chain can be factorized as the product of transfer matrices of two staggered XXZ spin chains with
fixed spectral difference
\bea t(u)=2^{8N}\rho_s(2u+i\pi-2\eta)\,\tilde t_s(u+i\pi)\, \tilde t_s(u), \label{20211223-1} \eea
where $\tilde t_s(u)= {\tilde t}(u)|_{\{\theta_j\}=\{0,\, i\pi\}}$.
The proof is as follows. For simplicity, we denote
\begin{equation}
\tilde T^s_{0'}(u)= \tilde T_{0'}(u)|_{\{\theta_j\}=\{0,\, i\pi\}}, \quad
\hat{\tilde T}^s_{0'} (u)=\hat{\tilde T}_{0'} (u)|_{\{\theta_j\}=\{0,\, i\pi\}}.
\end{equation}
From the direct calculation, we have \bea
\hspace{-0.8truecm}&&\hspace{-0.8truecm}\tilde t_s(u+i\pi)\tilde t_s(u)=[\rho_s(2u+i\pi-2\eta)]^{-1}tr_{0'0''}\{\tilde{K}^+_{0''}(u)\tilde{M}_{0''}^{-1}
\tilde R_{0'0''}(-2u+4\eta-i\pi)\tilde{M}_{0''} \no\\
\hspace{-0.8truecm}&&\hspace{-0.8truecm}\quad\quad\times \tilde{K}^+_{0'}(u)\tilde T^s_{0'} (u+i\pi)
\tilde T^s_{0''}(u)\tilde K_{0'}^-(u+i\pi)\tilde R_{0''0'}(2u+i\pi)\tilde K^-_{0''}(u)\hat{\tilde T}^s_{0'}
(u+i\pi)\hat{\tilde T}^s_{0''}(u)\}. \label{tt-1} \eea
By using the Yang-Baxter equation \eqref{Ttq11we11}, we obtain
\bea
\hspace{-1.2truecm}&& \hspace{-1.2truecm}\tilde R_{l'0'}(u+2i\pi)\tilde R_{l'0''}(u+i\pi)\tilde R_{0'0''}(-i\pi)\tilde R_{0''0'}(i\pi)
\tilde R_{j'0'}(u+i\pi)\tilde R_{j'0''}(u)\no\\
\hspace{-1.2truecm}&&\hspace{-1.2truecm}\qquad\qquad=\tilde R_{0'0''}(-i\pi)\tilde R_{l'0''}(u+i\pi)\tilde R_{l'0'}(u+2i\pi)\tilde R_{j'0''}(u)
\tilde R_{j'0'}(u+i\pi)\tilde R_{0''0'}(i\pi), \eea
which gives the identity
\bea\hat{\tilde T}^s_{0'}
(u+i\pi)\hat{\tilde T}^s_{0''}(u)=[\rho_s(i\pi)]^{-1}\tilde R_{0'0''}(-i\pi)\hat{\tilde T}^s_{0''}(u)\hat{\tilde T}^s_{0'}
(u+i\pi)\tilde R_{0''0'}(i\pi). \label{20211221-1}\eea
Substituting Eq.\eqref{20211221-1} into \eqref{tt-1}, we have
\bea
\hspace{-1.8truecm}&&\hspace{-1.8truecm}
\tilde t_s(u+i\pi)\,\tilde t_s(u)=[\rho_s(2u+i\pi-2\eta)\rho_s(i\pi)]^{-1}tr_{0'0''}
\{\tilde R_{0''0'}(i\pi)\tilde K^+_{0''}(u)\tilde {M}_{0''}^{-1}\no\\
\hspace{-1.0truecm}&&\hspace{-1.0truecm}\quad\quad\qquad\qquad\times \tilde R_{0'0''}(-2u+4\eta-i\pi)
\tilde{M}_{0''}{K}^+_{0'}(u)\tilde T^s_{0'} (u+i\pi)
\tilde T^s_{0''}(u)\no\\
\hspace{-1.0truecm}&&\hspace{-1.0truecm}\quad\quad\qquad\qquad\times
\tilde K^-_{0'}(u+i\pi)\tilde R_{0''0'}(2u+i\pi)\tilde K^-_{0''}(u)\tilde R_{0'0''}(-i\pi)\hat{\tilde T}^s_{0''}
(u)\hat{\tilde T}^s_{0'}(u+i\pi)\}\no\\
\hspace{-1.0truecm}&&\hspace{-1.0truecm}\quad\qquad\qquad=2^{-8N}[\rho_s(2u+i\pi-2\eta)]^{-1}{\cal{S}}^{-1}t(u){\cal{S}},\no
\eea
where ${\cal{S}}=S\otimes S\otimes\ldots\otimes S$. Then we arrive at the conclusion \eqref{20211223-1}.

\section{Exact solution}
\setcounter{equation}{0}

Now, we derive the eigenvalue of transfer matrix $t(u)$ of the $D^{(2)}_2$ spin chain based on the factorization identity \eqref{20211223-1}.
According to Eq.\eqref{120211224-1}, we know that $\tilde t_s(u+i\pi)$ and $\tilde t_s(u)$ have common eigenstates.
Acting Eq.\eqref{20211223-1} on a common eigenstate, we obtain
\bea\Lambda(u)=2^{8N}\rho_s(2u+i\pi-2\eta)\,\tilde \Lambda_s(u+i\pi)\, \tilde \Lambda_s(u), \label{ta}\eea
where $\Lambda(u)$, $\tilde \Lambda_s(u+i\pi)$ and $\tilde \Lambda_s(u)$ are the eigenvalues of the transfer matrices
$t(u)$, $\tilde t_s(u+i\pi)$ and $\tilde t_s(u)$, respectively.

In order to obtain the eigenvalue of transfer matrix $\tilde
t_s(u)$ of the staggered XXZ spin chain, we should diagonalize the
transfer matrix $\tilde t(u)$ of the inhomogeneous XXZ spin chain
first. The method is fusion \cite{f1,f2,f4,f5,f6-1,f6}. The
main idea of fusion is that the $R$-matrix at the some special
points can degenerate into the projector operators. For the
present case, at the point of $u=2\eta$, the $R$-matrix
\eqref{Rs-matrix} degenerates into \bea \tilde
R_{1'2'}(2\eta)=P^{(1) }_{1'2'}S_{1'2'}^{(1)},\eea where
$S_{1'2'}^{(1)}$  is an irrelevant constant matrix omitted here,
$P^{(1)}_{1'2'}$ is the one-dimensional projector operator \bea
P^{(1) }_{1'2'}=|\psi_0\rangle\langle\psi_0|, \quad
|\psi_0\rangle=\frac{1}{\sqrt{2\cosh\eta}} (e^{-\frac\eta
2}|12\rangle+e^{\frac\eta 2}|21\rangle), \eea and $\{|1\rangle,
|2\rangle \}$ are the orthogonal bases of the $2$-dimensional
linear space $V_{1'}$ (or $V_{2'}$). From the Yang-Baxter equation
\eqref{Ttq11we11} and using the properties of projector, we obtain
\bea
&&P^{(1) }_{2'1'}\tilde R _{1'3'}(u)\tilde R_{2'3'}(u+2\eta)P^{(1)}_{2'1'}=-\sinh(\frac u2+\eta)\sinh(\frac u2-\eta),\no \\
&&P^{(1) }_{1'2'}\tilde R _{3'1'}(u)\tilde R_{3'2'}(u+2\eta)P^{(1) }_{1'2'}=-\sinh(\frac{u}{2}+\eta)\sinh(\frac {u}{2}-\eta). \label{srf-1}
 \eea
Based on the reflections \eqref{rs12}-\eqref{rs1}, the fusion of reflection matrices gives
\bea
\hspace{-1.2truecm}&&\hspace{-1.2truecm} P_{1'2'}^{(1)}\tilde {K}^+_{2'}(u+2\eta)\tilde M_{1'} \tilde R_{1'2'}(-2u+2\eta)\tilde M_{1'}^{-1}
\tilde {K}^+_{1'}(u)P_{2'1'}^{\rm(1)}\no\\[4pt]
\hspace{-1.2truecm}&&\hspace{-1.2truecm}\qquad\qquad=2\sinh(u-2\eta)\frac{1}{\alpha'}\cosh\frac{u+\alpha'_1}{2}\cosh\frac{u-\alpha'_1}{2}
\cosh\frac{u+\alpha'_2}{2}\cosh\frac{u-\alpha'_2}{2}, \\[4pt]
\hspace{-1.2truecm}&&\hspace{-1.2truecm} P_{2'1'}^{(1)}\tilde K_{1'}^-(u)\tilde R_{2'1'}(2u+2\eta)\tilde K^-_{2'}(u+2\eta)P_{1'2'}^{(1)}\no\\[4pt]
\hspace{-1.2truecm}&&\hspace{-1.2truecm}\qquad\qquad=-2\sinh(u+2\eta)\frac{1}{\alpha}\cosh\frac{u+\alpha_1}{2}\cosh\frac{u-\alpha_1}{2}
\cosh\frac{u+\alpha_2}{2}\cosh\frac{u-\alpha_2}{2},\label{skf-2}
\eea
where the related constants are defined as
\bea &&\alpha=\frac{1}{2s_1s_2},\quad
\beta=\sqrt{\frac{8s_1s_2\cosh(2s)+16(s_1s_2)^2+1}{16(s_1s_2)^2}}, \quad \cosh\alpha_1=\frac{\alpha}{2}+\beta, \no\\[6pt]
&&
\cosh\alpha_2=\frac{\alpha}{2}-\beta, \quad \alpha'=\frac{1}{2s'_1s'_2},\quad
\beta'=\sqrt{\frac{8s'_1s'_2\cosh(2s')+16(s'_1s'_2)^2+1}{16(s'_1s'_2)^2}},\no\\[6pt]
&&\cosh\alpha'_1=\frac{\alpha'}{2}+\beta',\
\cosh\alpha'_2=\frac{\alpha'}{2}-\beta'. \no \eea
The Yang-Baxter relations \eqref{Ttqwe11} at certain points give
\bea
&&{\tilde T}_{0'}(\theta_j){\tilde T}_{0''}(\theta_j+2\eta)=P^{(1)}_{0''0'}{\tilde T}_{0'}(\theta_j)
{\tilde T}_{0''}(\theta_j+2\eta),\no\\[4pt]
&&\hat{\tilde T}_{0'}(-\theta_j)\hat{\tilde T}_{0''}(-\theta_j+2\eta)=P^{(1)}_{0'0''}
\hat{\tilde T}_{0'}(-\theta_j)\hat{\tilde T}_{0''}(-\theta_j+2\eta),
\label{stht-1}
 \eea
which show two ways to generate the projector operator in the transfer matrix.

Considering the physical quantity $\tilde t(\pm\theta_j)\tilde t(\pm\theta_j+2\eta)$ and using the fusion relations (\ref{srf-1})-(\ref{stht-1}), we obtain
\bea
\hspace{-1.truecm}&&\hspace{-1.truecm}
\tilde t(\pm\theta_j)\,\tilde t(\pm\theta_j+2\eta)=\frac{4\sinh(\pm{\theta_j}-2\eta)\sinh(\pm{\theta_j}+2\eta)}
{\alpha\alpha'\sinh(\pm\theta_j-\eta)\sinh(\pm\theta_j+\eta)}
\cosh\frac{\pm\theta_j-\alpha_1}{2}
\no\\[4pt]
\hspace{-1.truecm}&&\hspace{-1.truecm}\qquad\qquad\times\cosh\frac{\pm\theta_j+\alpha_1}{2}
\cosh\frac{\pm\theta_j-\alpha_2}{2}\cosh\frac{\pm\theta_j+\alpha_2}{2}
\cosh\frac{\pm\theta_j-\alpha'_1}{2}\cosh\frac{\pm\theta_j+\alpha'_1}{2}
\no\\[4pt]
\hspace{-1.truecm}&&\hspace{-1.truecm}\qquad\qquad\times
\cosh\frac{\pm\theta_j-\alpha'_2}{2}\cosh\frac{\pm\theta_j+\alpha'_2}{2}
\prod_{i=1}^M\sinh\frac{\pm\theta_j-\theta_i-2\eta}{2}\no\\[4pt]
\hspace{-1.truecm}&&\hspace{-1.truecm}\qquad\qquad\times\sinh\frac{\pm\theta_j-\theta_i+2\eta}{2}
\sinh\frac{\pm\theta_j+\theta_i-2\eta}{2}\sinh\frac{\pm\theta_j+\theta_i+2\eta}{2}, \;\; j=1, \cdots, 2N.\label{sottf1}
\eea
We see that the product of two transfer matrices with fixed spectral parameters is a c-number equaling to the quantum determinant
at the point of $u=\theta_j$. We shall note that the fusion identities \eqref{sottf1} hold only at the discrete inhomogeneous points. Besides, from the direct calculation and using the properties \eqref{Crossing-Unitarity-1},
we also obtain the values of $\tilde t(u)$ at the points of $u=0, 2\eta, i\pi$ as
\bea &&\tilde t(0)=\tilde t(2\eta)=2\cosh\eta\sinh s\sinh
s'\prod_{j=1}^{2N}\rho_s(\theta_j),\no\\
&&\tilde t(i\pi)=2\cosh\eta\cosh s\cosh
s'\prod_{j=1}^{2N}\rho_s(\theta_j+i\pi). \label{sottf2}\eea
The asymptotic behavior of $\tilde t(u)$ when the spectral parameter tends to infinity reads \bea\tilde  t(u)|_{u\rightarrow
\pm\infty}=-2^{-4N-2}e^{\pm[(2N+2)(u-\eta)]}(e^{-\eta}s_1s'_2+e^{\eta}s_2s'_1).\label{sottf3}\eea

From the definition \eqref{ts-1}, we know that the transfer matrix $\tilde t(u)$ is an operator polynomial of $e^{u}$ with the degree $4N+4$,
which can be completely determined by $4N+5$ constraints. Thus the above $4N$ fusion identities \eqref{sottf1} and $5$ additional conditions
\eqref{sottf2}-\eqref{sottf3} give us sufficient information to determine the eigenvalue $\tilde \Lambda(u)$ of $\tilde t(u)$.
After some algebras, we express the eigenvalue $\tilde \Lambda(u)$ as the inhomogeneous $T-Q$ relation
\bea
&&\hspace{-0.2cm}\tilde  \Lambda(u)=\frac{2\sinh(u-2\eta)}
{\sinh(u-\eta)\sqrt{\alpha\alpha'}}
\cosh\frac{u+\alpha_1}{2}\cosh\frac{u+\alpha_2}{2}\cosh\frac{u+\alpha'_1}{2}\cosh\frac{u+\alpha'_2}{2}
 a(u)\frac{Q(u+2\eta)}{Q(u)}\no\\
&&\qquad\quad+\frac{2\sinh u}
{\sinh(u-\eta)\sqrt{\alpha\alpha'}}\cosh\frac{u-2\eta-\alpha_1}{2}\cosh\frac{u-2\eta-\alpha_2}{2}\cosh\frac{u-2\eta-\alpha'_1}{2}\no\\
&&\qquad\quad\times
\cosh\frac{u-2\eta-\alpha'_2}{2}d(u)\frac{Q(u-2\eta)}{Q(u)}+x\sinh u\sinh(u-2\eta)\frac{a(u)d(u)}{Q(u)},\label{tse}\eea
where the functions $Q(u)$, $a(u)$, $d(u)$ and parameter $x$ are
\bea &&\hspace{-0.3cm}Q(u)=\prod_{l=1}^{2N}\sinh\frac 12(u-\mu_l)\sinh\frac
12(u+\mu_l-2\eta), \no\\
&&\hspace{-0.3cm}a(u)=\prod_{j=1}^{2N}\sinh\frac12(u-\theta_j-2\eta)
\sinh\frac12(u+\theta_j-2\eta)=d(u-2\eta),\no \\
&&\hspace{-0.3cm}x=-2\sqrt{s_1s_2s'_1s'_2}\cosh[(2N+1)\eta+\frac{\alpha_1+\alpha_2+\alpha'_1+\alpha'_2}{2}]-(e^{-\eta}s_1s'_2+e^{\eta}s_2s'_1).\eea
Because $\tilde  \Lambda(u)$ is a polynomial, the singularities of right hand side of Eq.\eqref{tse} should be cancelled with each other, which gives that the
Bethe roots $\{\mu_l\}$ should satisfy the Bethe ansatz equations
\bea
\hspace{-0.6truecm}&&\hspace{-0.6truecm}\frac{2\sinh(\mu_l-2\eta)}
{\sinh(\mu_l-\eta)\sqrt{\alpha\alpha'}}
\cosh\frac{\mu_l+\alpha_1}{2}\cosh\frac{\mu_l+\alpha_2}{2}\cosh \frac{\mu_l+\alpha'_1}{2} \cosh \frac{\mu_l+\alpha'_2}{2}\frac{Q(\mu_l+2\eta)}{d(\mu_l)}\no\\
\hspace{-0.6truecm}&&\hspace{-0.6truecm}\quad+\frac{2\sinh \mu_l}
{\sinh(\mu_l-\eta)\sqrt{\alpha\alpha'}}\cosh\frac{\mu_l-2\eta-\alpha_1}{2}
\cosh\frac{\mu_l-2\eta-\alpha_2}{2}\cosh\frac{\mu_l-2\eta-\alpha'_1}{2}\no\\
\hspace{-0.6truecm}&&\hspace{-0.6truecm}\quad\quad\times \cosh\frac{\mu_l-2\eta-\alpha'_2}{2}
\frac{Q(\mu_l-2\eta)}{a(\mu_l)}=-x\sinh \mu_l\sinh(\mu_l-2\eta), \quad l=1, \cdots, 2N. \label{BA0}\eea

Some remarks are in order. By solving the algebraic equations (\ref{BA0}), we obtain the values of Bethe roots $\{\mu_l\}$.
Substituting these values into the inhomogeneous $T-Q$ relation (\ref{tse}), we obtain the eigenvalue $\tilde \Lambda(u)$.
The different sets of Bethe roots would give different eigenvalues. As shown in \cite{Cao14, Cao15}, based on the numerical calculation
and analytical analysis with the help of B\'{e}zout theorem, the $T-Q$ relation (\ref{tse}) can generate all the eigenvalues of $\tilde t(u)$.
The eigenvalue $\tilde \Lambda(u)$ has the well-defined quasi-inhomogeneous limit $\{\theta_j\}=\{0,\, i\pi\}$. Substituting
\bea
\tilde \Lambda_s(u)=\tilde \Lambda(u)|_{\{\theta_j\}=\{0,\, i\pi\}}, \quad \tilde \Lambda_s(u+i\pi)=\tilde \Lambda(u+i\pi)|_{\{\theta_j\}=\{0,\, i\pi\}},
\eea
into Eq.\eqref{ta}, we then are able to obtain the eigenvalue $\Lambda(u)$ of the transfer matrix $t(u)$ of the $D^{(2)}_2$ spin chain associated with the most generic non-diagonal $K$-matrices $K^{\pm}(u)$ given by (\ref{kM})-(\ref{kd21q3-1}) . Therefore,
the expression (\ref{ta}) gives the complete spectrum of the system via the relation (\ref{hh}).

\section{Discussion}

In this paper, we have studied the exact solutions of one-dimensional quantum integrable system connected with the twisted $D^{(2)}_2$ quantum algebra in the generic open boundary conditions,
where the reflection matrices have non-diagonal elements. We find that the generating functional of the model can be factorized as the product of transfer matrices of two XXZ spin chains with
staggered inhomogeneous parameters. Based on these factorization identities and using the method of fusion,
we obtain the eigenvalues and corresponding Bethe ansatz equations of the model.

Based on the obtained eigenvalues, the eigenstate of the $D^{(2)}_2$ model can be retrieved by using the separation of variables \cite{Skl95,Fra08, Fra11, Nic12} or the off-diagonal Bethe ansatz \cite{Zhan15-1}.
Then the correlation functions, norm, form factors and other interesting scalar products can be calculated.
Staring from the obtained Bethe ansatz equations and using the finite size scaling analysis of the contribution of inhomogeneous term in the $T-Q$ relation (\ref{tse}),
the physical quantities such as ground state energy density, surface energy and elementary excitations in the thermodynamic limit could also be studied.
The results given in this paper are the foundations to exactly solve the high rank $D^{(2)}_{n}$ model by using the analytical methods such as the nested off-diagonal Bethe ansatz \cite{wang15}.

\section*{Acknowledgments}

The financial supports from National Key R\&D Program of China (Grant No. 2021YFA1402104),
the National Natural Science Foundation of China (Grant Nos. 12074410, 12047502, 12075177, 11934015,
11975183, 11947301 and
91536115), Major Basic Research Program of Natural Science of
Shaanxi Province (Grant Nos. 2021JCW-19, 2017ZDJC-32), Australian
Research Council (Grant No. DP 190101529), Strategic
Priority Research Program of the Chinese Academy of Sciences (Grant No. XDB33000000), Shaanxi Province Key Laboratory
of Quantum Information and Quantum Optoelectronic Devices, Xi'an
Jiaotong University, and Double First-Class University Construction Project of Northwest
University are gratefully acknowledged.

\section*{Appendix A. Another non-diagonal boundary reflection}
\setcounter{equation}{0}
\renewcommand{\theequation}{A.\arabic{equation}}
The reflection equation \eqref{r1} has another inequivalent generic non-diagonal solution where the matrix elements are
\bea&&k_{11}(u)=\frac{1}{2}e^{-u}[\sinh(u-\eta)\cosh(u-2s)-2s_1s_2\sinh\eta\cosh^2 u],\no\\
&&k_{12}(u)=-\frac{1}{2}s_1e^{-\frac u2+\frac{
i\pi}{4}}\sqrt{\cosh\eta}\sinh(2u)\sinh\frac
12(u-\eta-2s+\frac{i\pi}{2}),\no\\
&&k_{13}(u)=\frac{1}{2}s_1e^{-\frac
u2+\frac{i\pi}{4}}\sqrt{\cosh\eta}\sinh(2u)\cosh\frac
12(u-\eta-2s+\frac{i\pi}{2}),\no\\
&&k_{14}(u)=-\frac{1}{2}s_1^2\cosh u\sinh(2u),\no\\
&&k_{21}(u)=\frac{1}{2}s_2e^{-\frac
u2+\frac{i\pi}{4}}\sqrt{\cosh\eta}\sinh(2u)\cosh\frac
12(u-\eta-2s+\frac{i\pi}{2}),\no\\
&&k_{22}(u)=-\frac{1}{2}\cosh u[\sinh\eta\cosh(2s)-2s_1s_2\cosh u\sinh(u-\eta)],\no\\
&&k_{23}(u)=\frac{1}{2}\sinh u [\cosh\eta\sinh(2s)-\sinh(u+\frac{i\pi}{2})],\no\\
&&k_{24}(u)=\frac{1}{2}s_1e^{\frac
u2+\frac{i\pi}{4}}\sqrt{\cosh\eta}\sinh(2u)\cosh\frac
12(u-\eta+2s-\frac{i\pi}{2}),\no\\
&&k_{31}(u)=-\frac{1}{2}s_2e^{-\frac
u2+\frac{i\pi}{4}}\sqrt{\cosh\eta}\sinh(2u)\sinh\frac
12(u-\eta-2s+\frac{i\pi}{2}),\no\\
&&k_{32}(u)=-\frac{1}{2}\sinh u[\cosh\eta\sinh(2s)-\sinh(u+\frac{i\pi}{2})],\no\\
&&k_{33}(u)=-\frac{1}{2}\cosh u[\sinh\eta\cosh(2s)-2s_1s_2\cosh u\sinh(u-\eta)],\no\\
&&k_{34}(u)=\frac{1}{2}s_1e^{\frac
u2+\frac{i\pi}{4}}\sqrt{\cosh\eta}\sinh(2u)\sinh\frac
12(u-\eta+2s-\frac{i\pi}{2}),\no\\
&&k_{41}(u)=-\frac{1}{2}s_2^2\cosh u\sinh(2u),\no\\
&&k_{42}(u)=\frac{1}{2}s_2e^{\frac u2+\frac{i\pi}{4}}\sqrt{\cosh\eta}\sinh(2u)\sinh\frac
12(u-\eta+2s-\frac{i\pi}{2}),\no\\
&&k_{43}(u)=\frac{1}{2}s_2e^{\frac u2+\frac{i\pi}{4}}\sqrt{\cosh\eta}\sinh(2u)\cosh\frac
12(u-\eta+2s-\frac{i\pi}{2}),\no\\
&&k_{44}(u)=\frac{1}{2}e^{u}[\sinh(u-\eta)\cosh(u+2s)-2s_1s_2\sinh\eta\cosh^2 u].\label{sskd-1}
\eea
In this case, the reflection matrices ${K}^{\pm}_1(u)$ of the $D^{(2)}_2$ spin chain can be also factorized as the product of
reflection matrices $\tilde {K}^{\pm}_{1',2'}(u)$ of the XXZ spin chain by a different way from those of \eqref{kd1}
\bea
&&{K}^{+}_1(u)=S{\cal \tilde P}_{1'2'}\tilde {K}^+_{2'}(u+\frac{i\pi}{2})\tilde{M}_{2'}^{-1}
\tilde R_{1'2'}(-2u+4\eta-2i\pi) \tilde{M}_{2'}\tilde{K}^+_{1'}(u+\frac{3i\pi}{2})S^{-1}, \no \\
&&K^{-}_1(u)=S\tilde K_{1'}^-(u+\frac{3i\pi}{2})\tilde
R_{2'1'}(2u+2i\pi)\tilde K^-_{2'}(u+\frac{i\pi}{2}){\cal\tilde
P}_{1'2'}S^{-1}, \label{kd11} \eea where the permutation operator
${\cal \tilde P}_{1'2'}$ is included. Here it should  be addressed
that when $\tilde K^-(u)=1$ in (\ref{kewesk111}), the resulting $K^-(u)$
given by (\ref{kd11}) is that discussed in \cite{b5} with
$\epsilon=1$. When $s_1=s_2=0$ in (\ref{kewesk111}), the resulting $K^-(u)$
given by (\ref{kd11}) is the third case  discussed in
\cite{26}. For $K^-(u)$ defined by (\ref{sskd-1}), the corresponding
$K^+(u)$ given by (\ref{kM})  indeed satisfies $tr_0
K^{+}_0(0)= 0$. For this case one has to, instead of (\ref{hh}), take  the second order derivative of the
transfer matrix to construct a meaningful Hamiltonian\cite{b1}.

Motivated by the factorization \eqref{kd11}, we construct the transfer matrix of the related XXZ spin chain as
\begin{equation}
\bar{t}(u)= tr_{0'} \{\tilde K^+_{0'}(u+\frac{i\pi}{2})\tilde T_{0'} (u)
\tilde K^-_{0'}(u+\frac{i\pi}{2})\hat{\tilde T}_{0'} (u+{i\pi})\}. \label{ts-2}
\end{equation}
The transfer matrix $\bar{t}(u)$ can be obtained from $\tilde t(u)$ (\ref{ts-1}) by the mapping
\bea
\bar{t}(u)=\tilde t(u)|_{u\rightarrow
u+\frac{i\pi}{2},\,\{\theta_j\rightarrow \theta_j+\frac{i\pi}{2}\}}.
\eea

After some algebraic calculation, we find that if the inhomogeneous parameters in Eq.\eqref{ts-2} are staggered, i.e., $\theta_j=\frac{i\pi}{2}$ for odd $j$ and $\theta_j=\frac{3i\pi}{2}$ for even $j$,
the transfer matrix $t(u)$ of $D^{(2)}_2$ spin chain can be factorized as the product of transfer matrices of two staggered XXZ spin chains with
fixed spectral difference
\bea t(u)=2^{8N}\rho_s(2u+2i\pi-2\eta)\,\bar{t}_s(u+i\pi)\,\bar{t}_s(u), \label{1}\eea
where
\begin{equation}
\bar{t}_s(u)=\bar{t}(u)|_{\{\theta_j\}=\{i\pi/2,\, 3i\pi/2\}}.\label{2}
\end{equation}
The proof is as follows. From the definition \eqref{2}, we readily have
\bea
\hspace{-0.8truecm}&&\hspace{-0.8truecm}
\bar{t}_s(u+i\pi)\,\bar{t}_s(u)=[\rho_s(2u+2i\pi-2\eta)]^{-1}tr_{0'0''}\{\tilde {K}^+_{0''}(u+\frac{i\pi}{2})\tilde{M}_{0''}^{-1}
 \no\\
\hspace{-0.8truecm}&&\hspace{-0.8truecm}\qquad\qquad\times \tilde R_{0'0''}(-2u+4\eta-2i\pi) \tilde{M}_{0''}\tilde{K}_{0'}(u+\frac{3i\pi}{2})\tilde T^s_{0'}(u+i\pi)
\tilde T_{0''}^s(u)\tilde K^-_{0'}(u+\frac{3i\pi}{2})\no\\
\hspace{-0.8truecm}&&\hspace{-0.8truecm}\qquad\qquad\times\tilde R_{0''0'}(2u+2i\pi)\tilde K^-_{0''}(u+\frac{i\pi}{2})\hat{\tilde T}^s_{0'}
(u+2i\pi)\hat{\tilde T}^s_{0''}(u+i\pi)\}. \label{tt-2} \eea
By using the property ${\cal \tilde P}_{0''0'}^2=1$, we obtain
\bea
\hspace{-1.2truecm}&&\hspace{-1.2truecm}\tilde R_{j'0'}(u+3i\pi)\tilde R_{l'0'}(u+2i\pi)
\tilde R_{j'0''}(u+2i\pi)\tilde R_{l'0''}(u+i\pi)\no\\
\hspace{-1.2truecm}&&\hspace{-1.2truecm}\qquad\qquad={\cal \tilde P}_{0''0'}\tilde R_{j'0''}(u+3i\pi)\tilde R_{l'0''}(u+2i\pi)
\tilde R_{j'0'}(u+2i\pi)\tilde R_{l'0'}(u+i\pi){\cal \tilde P}_{0''0'}, \eea
which gives
\bea
\hat{\tilde T}_{0'}^s
(u+2i\pi)\,\hat{\tilde T}^s_{0''}(u+i\pi)={\cal\tilde  P}_{0'0''}\,\hat{\tilde T}^s_{0''}(u)\,\hat{\tilde T}^s_{0'}
(u+i\pi)\,{\cal \tilde  P}_{0'0''}. \label{3}\eea
Substituting Eq.\eqref{3} into \eqref{tt-2}, we obtain
 \bea
\bar{t}_s(u+i\pi)\,\bar{t}_s(u)=2^{-8N}[\rho_s(2u+2i\pi-2\eta)]^{-1}\,{\cal{S}}^{-1}\,t(u)\,{\cal{S}},\label{ttt-2}
\eea
which gives the conclusion \eqref{1}.

Because the transfer matrices $\bar{t}(u)$ with different spectral parameters commutate with each other, they have the common eigenstates.
Acting the factorization identity \eqref{1} on a common eigenstate, we obtain the eigenvalue $\Lambda(u)$ of the transfer matrix $t(u)$
of the $D^{(2)}_2$ spin chain as
\bea\Lambda(u)=2^{8N}\rho_s(2u+2i\pi-2\eta)\,\bar{\Lambda}_s(u+i\pi)\,\bar{\Lambda}_s(u), \eea
where
\bea
\bar{\Lambda}_s(u)=\tilde{\Lambda}(u)|_{u\rightarrow u+\frac{i\pi}{2}, \{\theta_j\}=\{i\pi/2,\, 3i\pi/2\}, \, \{\mu_l\rightarrow
\mu_l+\frac{i\pi}{2}\}}, \eea
and $\tilde{\Lambda}(u)$ is given by Eq.(\ref{tse}).

\end{document}